\documentclass{article}

\usepackage{arxiv}

\usepackage[utf8]{inputenc} 
\usepackage[T1]{fontenc}    
\usepackage{hyperref}       
\usepackage{url}            
\usepackage{booktabs}       
\usepackage{amsfonts}       
\usepackage{nicefrac}       
\usepackage{microtype}      
\usepackage{lipsum}
\usepackage{graphicx}
\graphicspath{ {./images/} }
\usepackage{framed,color,verbatim}
\definecolor{shadecolor}{rgb}{.9, .9, .9}
\newenvironment{code}%
   {\snugshade\verbatim}%
   {\endverbatim\endsnugshade}

\title{Mitigating Docker Security Issues}

\author{
  Robail Yasrab \\
  Institute of Biomedical Engineering, Department of Engineering Science,\\ University of Oxford, OX3-7DQ, Oxford. \\ University of Cambridge\\  
  \texttt{\{robail.yasrab@eng.ox.ac.uk, robail.yasrab@mrc-bsu.ac.uk\}} 

}

\begin{document}
\maketitle
\begin{abstract}

Docker offers an ecosystem that offers a platform for application packaging, distributing, and managing within containers. However, the Docker platform has not yet matured. Presently, Docker is less secured than virtual machines (VM) and most of the other cloud technologies. The key to Docker’s inadequate security protocols is; container sharing of Linux kernel, which can lead to the risk of privileged escalations. This research will outline some significant security vulnerabilities at Docker and counter solutions to neutralize such attacks. There are a variety of security attacks like insider and outsider. This research will outline both types of attacks and their mitigations strategies. Taking some precautionary measures can save from massive disasters. This research will also present Docker secure deployment guidelines. These guidelines will suggest different configurations to deploy Docker containers in a more secure way. 
\end{abstract}


\section{INTRODUCTION}
As machine learning and artificial intelligence become increasingly prevalent in today's world [27-37], people have been doubtful regarding the security of cloud technologies since their inception. People do not trust cloud computing security aspects entirely as this is still an emerging phenomenon. The security vendors design a marketing response for clients by using standard terms to market their products [1]. Though many issues are lurking around, new cloud security solutions mitigate many security and privacy vulnerabilities. 

Currently, cloud technologies are transforming traditional technology with new and more efficient practices [2,3]. One such example is containers. Containers are considered the future of virtual machines (VM). Container-based virtualization is also known as operating system virtualization, which allows the virtualization layer to run as an application within the operating system (OS) [4]. Containers appeared as micro virtual machines, more lightweight and more efficient because there is just one operating system (OS) managing all hardware calls.

Presently, the container model is misrepresented. Container technology looks secure in that it contains all dependencies in one package. However, it does not ensure its security. Container platforms are also vulnerable as other cloud platforms. Several threats are threatening from inside and outside [5]. 
Virtual environments offer plenty of challenges to run services with outstanding security, particularly in a multi-tenant cloud system [6]. It is already established that virtual machines (VMs) developed through hypervisor-based virtualization methods are highly secured compared to containers. The key reason behind this aspect is that VMs add a layer of isolation among the host and the applications.

VM-based applications are only allowed to communicate with the VM kernel, not to the host kernel. So, an attacker has to bypass the VM kernel and the hypervisor before an attack can be made on the host kernel. On the other hand, the containers model offers an application to directly access and communicate with the host kernel, as shown in Fig. 1. Such a situation permits an attacker to hit the host kernel directly. This is one of the critical aspects that raise security concerns about container technology compared to VM platforms.   
\begin{figure}
\centerline{\includegraphics[width=17cm, height=10cm]{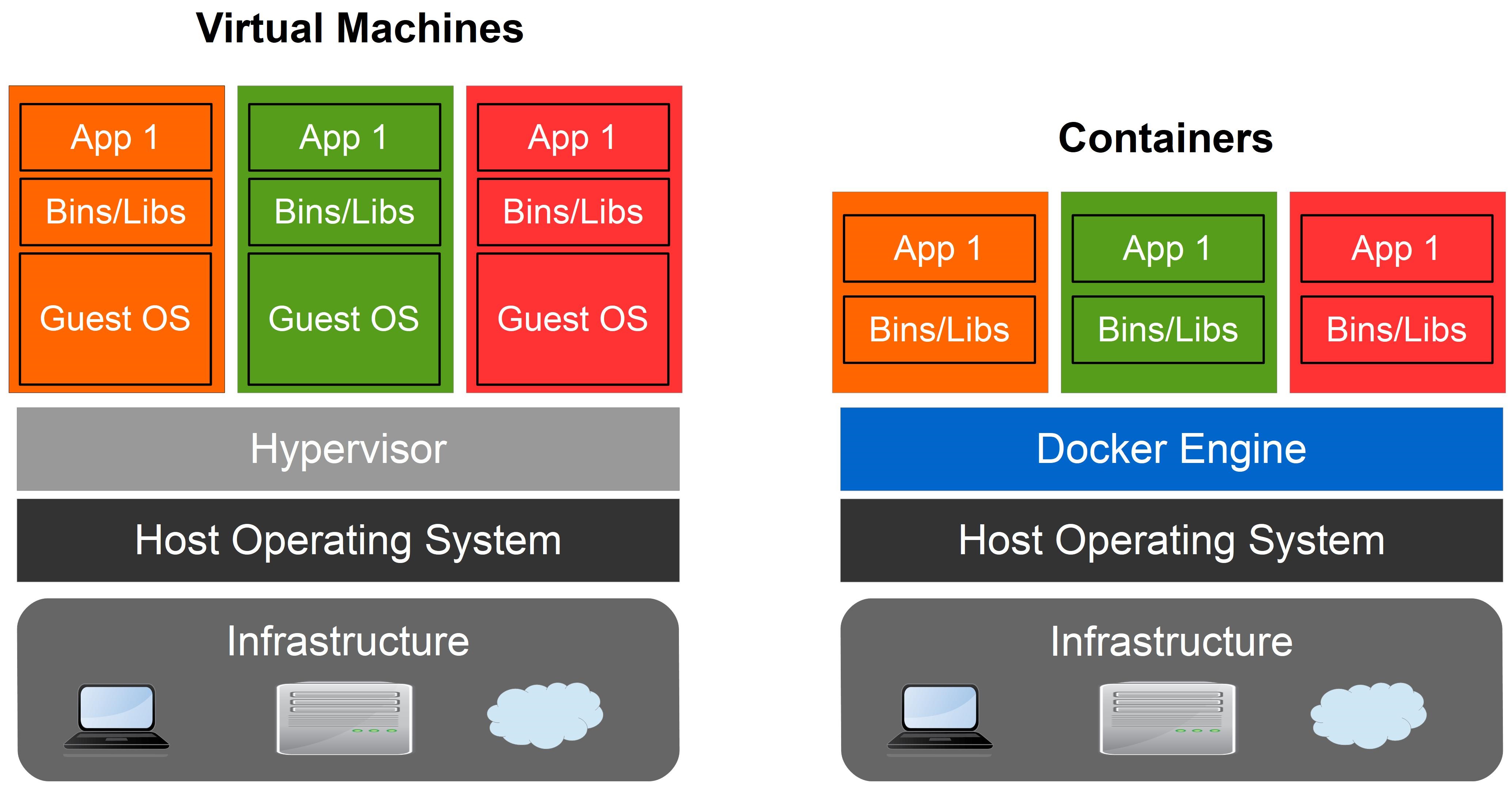}}
\caption{VMs vs. Container}
\label{fig1}
\end{figure}                 

Currently, Docker appeared as one of the top container-based virtualization platforms. Docker containers also suffer from the same security vulnerabilities. Docker demonstrates the idea of less friction; however, security is just the opposite of it.

The research is aimed to offer a deep insight into Docker and its analysis, which leads to the following research question. Is Docker offering a safe environment to run applications? The analysis will discover some of the critical security vulnerabilities which are currently threatening Docker. This research will also highlight the critical mitigation policies to avoid such issues. This analysis will attempt to assess the majority of security threats internally as well as externally. 
The initial section of this paper is about the detailed overview of the Docker platform. The second section will outline some of the significant security issues and their mitigation strategies. The third section will outline recommendations for the security deployment of containers at the Docker platform and a new evolutionary hybrid architecture that is more secure and efficient. The last section is based on the conclusion and recommendations.

\section{DOCKER OVERVIEW} 
Docker has quickly turned out as one of the leading projects for containerizing applications. Docker platform was initiated as an open-source project that allows packing, shipping, and running applications in lightweight containers. Docker containers offer unique capabilities of platform-agnostic and hardware-agnostic. These containers do not have any dependences regarding a particular framework, language, or packaging system. Docker containers can run within any technology-based environment. This capability makes these containers independent from a particular stack or provider [7]. 
Solomon Hykes initiated Docker as an open-source internal project at dotCloud, a platform-as-a-service (PaaS) company. The initial version of open-source Docker was released in March 2013. Docker was designed to utilize different interfaces to access virtualization characteristics of the Linux kernel, as shown in Fig.2. Initially, Docker was utilizing LXC as the default execution environment for its platform. However, on version 0.9 on March 13, 2014, Docker dropped LXC and introduced its libcontainer library. Go programming language was used to write Docker libcontainer library [8]. 

\begin{figure*}[h]
\centerline{\includegraphics[width=14cm, height=10cm]{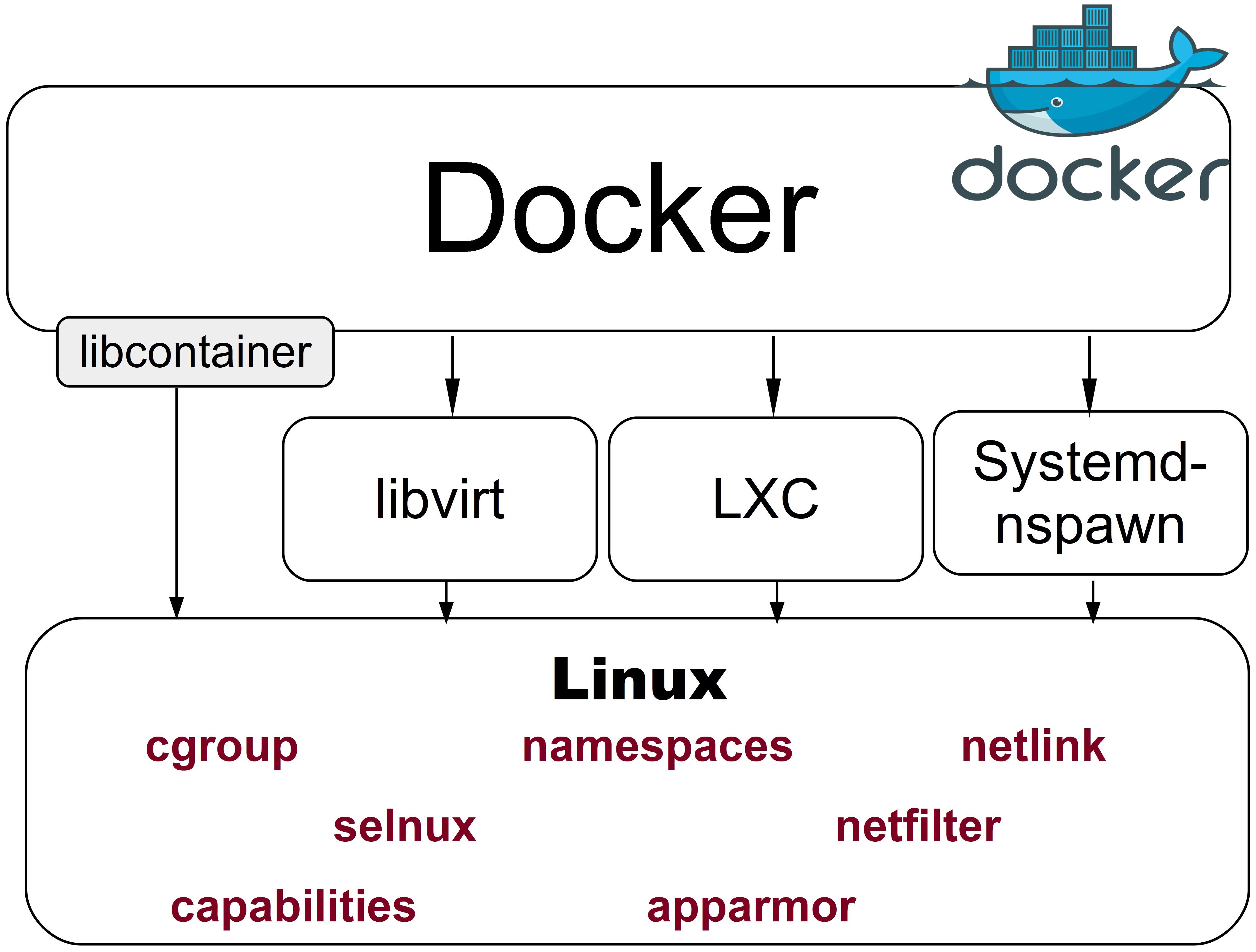}}
\caption{Docker uses virtualization features of the Linux kernel}
\label{fig1}
\end{figure*}                     

The process of developing containers on Docker is based on a number of steps. For developing containers, we search for images in the local Docker library. Docker offers images command lists locally to create a Docker container. However, if the required image is not available, it can be downloaded from the central Docker repository. The next step is to build a container as well as formulate necessary changes. The whole process is shown in Fig. 3.
\begin{figure*}[h]
\centerline{\includegraphics[width=14cm, height=8.5cm]{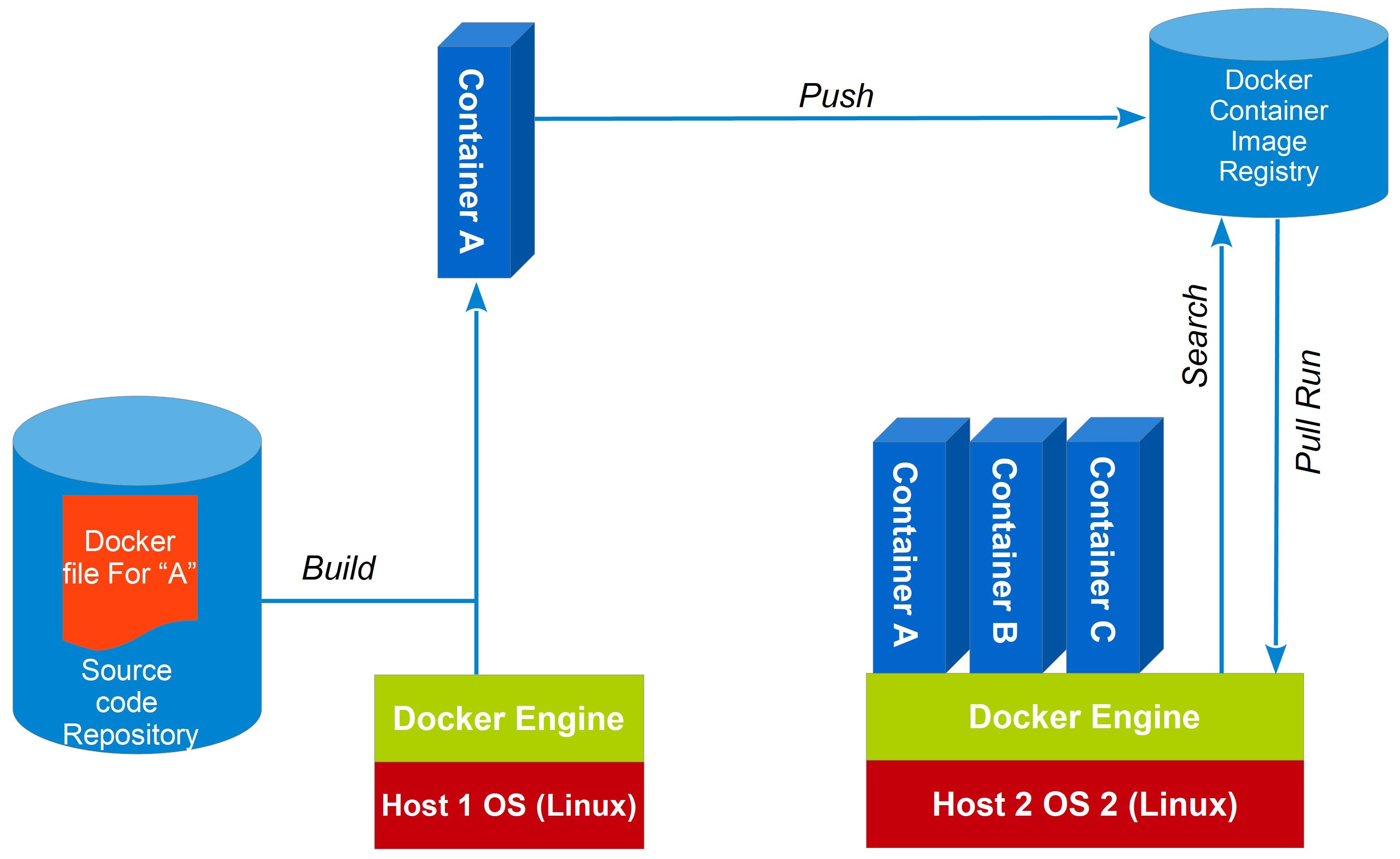}}
\caption{Developing a Docker container}
\label{fig1}
\end{figure*}     

Till now, several popular technology organizations worldwide showed their interest and dependence on Docker. Organizations like Red Hat, Google, IBM, Google, Amadeus IT Group, and Cisco Systems have already contributed to Docker at a great level. Red-Hat is one of the most significant contributors to Docker containers. Recently Red-Hat announced that users can now run Docker containers on Red Hat Enterprise Linux 7 (RHEL 7) and Red Hat Enterprise Linux Atomic (based on RHEL 7) systems [9]. 
On August 12, 2015, a new version of Docker 1.8 was announced. This new version offers new capabilities of content trust, toolbox, and updates to registry and orchestration. This new version will also offer support for image signing, a new installer, and incremental improvements to Engine, Swarm, Compose, Machine and Registry [10]. Docker 1.8 version offers support for users of Mac OS X, Linux, and Windows versions. 

\section{SECURITY ANALYSIS} 

The current growing technology market demands enhanced virtualization technologies. Though several virtualization methods are not flexible enough to satisfy developer needs. Most virtualization technology solutions offer considerable overhead that turns out to be a burden on the scalability of the infrastructure.

In such situations, Docker has evolved a lightweight virtualization solution that minimizes overhead by containerizing applications and services [11]. Docker offers the same kernel as the host system to minimize the resources overhead, though this technique can expose containers to several security risks if not effectively configured [12]. 

Running services in a multi-tenant cloud-based virtual environment presents security as one of the key challenges. VM allows an application to only communicate with the VM kernel, not with the host kernel. Conversely, container-based applications can directly communicate with the host kernel [13,14]. This is one of the critical weak points for containers security and privacy management. Docker is also based on a similar container-based virtualization methodology, also having the same security vulnerabilities.

To analyze the security vulnerabilities of Docker, we have to assess two sides of the system, 'outsider' and 'insider.' Security attacks can happen from outside and inside of the system. Outsider attacks try to get access to the container and damage the data and exploit resources. While insider attacks are carried out by a malicious user present inside, through getting access to the Docker commands. The fundamental objective of both kinds of attackers is identical, to damage and exploit container intellectual resources.
For example, a simple command (given below) can do massive damage if your system is not properly configured. The below-given command is executed through a dummy flag:
\begin{code}
docker run --dontpastethis --privileged -v /usr:/usr busybox rm -rf /usr
\end{code}

The above-stated command deletes all of the host’s /usr folder. So, we can assess that how much a system can be damaged if security precautions are not taken properly. 
This analysis aims to find out the most vulnerable security issues (insider/outsider) and suggest some precautions and mitigation strategies to counter such attacks.   
Insider attacks can be managed through some simple configurations. Running Docker Daemon with –selinux flag will prevent containers from doing damage to the host system through developing an additional security layer. Further details of different insider attacks are outlined in the coming sections. 

Let’s consider the security of the container in comparison to the security of a ship. The ship acts as a dock. It can build a powerful and secure container, and however, if the ship is insecure, it does not matter how strong the container is? Similarly, a great deal of care should be taken to avoid outsider attacks. By removing the capabilities of containers, several outsider security threats can be avoided. Container capabilities are the partition of roots into 32 different categories. By default, many of these capabilities are disabled in Docker containers. For instance, by default, the Docker container’s IPTables rules cannot be manipulated. For disabling all of these capabilities, the command is given below:
\begin{code}
docker run -ti --cap-drop ALL debian /bin/bash   
\end{code}

\section{DEFENDING AGAINST SECURITY ATTACKS }
Security is now one of the critical aspects of cloud application. Docker also suffers from several security attacks. This section will outline different insider and outsider security attacks and their mitigation policies. 

\subsection{Kernel Exploits}
Kernel mange and deals with all container operations and processes. In the case of a kernel-level exploit, the applications running inside containers are on the verge of compromising and exploit. All containers share the same kernel architecture [5]. In this situation, if some contained application is hijacked and obtain some privileged rights of the kernel, then such condition leads to compromise all running containers and host platform. Likewise, there is no possibility that two containers utilize different versions of the same kernel module.

Docker is currently taking security very seriously and offering more solid and practical solutions to tackle and deal with significant security issues. Docker recommends the following precautions to implement a secure and safe cloud environment to deal with kernel-level exploits.    
It is recommended that AppArmor or SELinux should be executed while running Docker Engine.

Docker states that mutually trusted containers should be mapped together in groups at separate machines. Untrusted applications should not be running with root privileges. Docker Engine has also started support for user namespaces, offering an additional layer of security for containers.
To avert kernel exploits, the container file system must be set to read-only. By turning off inter-container communication, such attacks can be avoided. Avoiding unnecessary package installations in the container is also an excellent way to keep such dangers away. 
\subsection{Denial of Service (Dos) Attacks}
Denial-of-Service or DoS is one of the most well-known attacks on network resources. In such attacks, a process or a group of processes try to consume the entire resources of the system, thus breaking down or disrupting standard processing or operations.
In containers based processing architecture, all containers share kernel resources. DoS condition happens when one container exploits access to a resource. In such conditions, it will starve out all other containers. 
To encounter such attacks, there is a need for OS-level virtualization solutions that should fulfill requirements like file system isolation, process isolation, IPC isolation, device isolation, network isolation, and controlling resource allocation [5].

So, by controlling resource allocation to each container, such attacks can be prevented. Docker implements Cgroups as a critical tool to deal with such issues. Cgroups control and manage the resource limits, e.g., CPU time, memory space, and disk I/O that any Docker container can use. They ensure that every container gets its fair share of the resources and avoiding any container monopolizes all resources. Moreover, Cgroups allow Docker to control and configure resource allocation constraints for every container. For instance, one such constraint is controlling the CPUs availability to a specific container [5].
\subsection{Container Breakouts}
In such attacks, an attacker breaks out a container, then he/she can access the host and other containers. After getting access, the attacker will be able to access files outside the container.
open\_by\_handle\_at() function allows the process to access files on a mounted filesystem through file\_handle structure. file\_handle structure utilizes inode numbers to distinguish files. To call this function needs CAP\_DAC\_READ\_SEARCH capability. A superuser inside a container has this capability by default. This allows an attacker to bypass simfs constraints and access the entire files on a primary filesystem comprising other VE’s residing on a similar filesystem. 

According to Docker’s website, container breakout issue and vulnerability was only present until Docker version 0.11. According to the new details, this vulnerability was fixed in Docker 0.12, which was ultimately turned out to be Docker 1.0.
To mitigate such kinds of security vulnerabilities, there is a need to set the container file system to read-only. Running containers with the privileged flag can be dangerous and can cause such kinds of security attacks. Setting containers volumes to read-only is an effective way to discourage container breakout.  

\subsection{Poisoned Images}
Container images may be injected through some virus or trojan-infected software. The problem of poisoned images also happens if someone is running outdated, known-vulnerable software versions.  
According to Docker, a downloaded image is “verified” by the system. This verification is solely based on the presence of a signed manifest. Though Docker never authenticates downloaded image checksum from the manifest. In such situations, an attacker could transmit any image together with a signed manifest. Such kind of security issues can lead to numerous serious vulnerabilities [15].
In Docker, images are downloaded from an HTTPS server. These images pass through an insecure streaming processing pipeline inside Docker daemon:
\begin{code}
[decompress] -> [tarsum] -> [unpack]
\end{code}

Inside Docker, this pipeline is practical, however, extremely unprotected. Therefore, there is a need that unauthenticated input steam should not be assessed before confirming its digital signature. 
Docker users need to be alert that the code used for downloading images is shockingly insecure. In this scenario, users should only download authenticated and trusted images. Another better option to manage such security issues is to block index.docker.io locally. Through this, a user will download and authenticate images manually before importing them to the Docker platform through Docker load. 

\subsection{Compromised Secrets}
Compromising business or personal secrets are the significant security dangers in container-based technology. In case of theft of API keys and database passwords, the overall system can be compromised. Docker allows users to run multiple containers at the same time. In case of a security breach, overall services and operations can be disrupted. So a lot more is needed to be done to protect database passwords and API keys. Such details must be kept secret to avoid any possible security breach. 

To further protect Docker from such attacks, there is a need to set the container file system to read-only. For sharing secrets, utilizing environment variables is not a good option. Running containers without privileged flags will also be considered a great help to avoid compromising security attacks.   
\subsection{Man-in-the-Middle (MitM)} 
In such attacks, a malicious actor inserts himself/herself into communication among two legitimate parties. Such attackers monitor, alter or steal valuable information which transmits among two parties.    
To avoid such attacks in containers, network isolation is the most significant aspect to prevent such network-based attacks. There is a need to configure containers so that they are incapable of manipulating or eavesdrop on the network traffic of the host or other containers [5].
In this scenario, OpenVPN (an open virtual private network) offers the best way to implement virtual private networks (VPNs) employing TLS (Transport Layer Security) encryption. OpenVPN defends the network traffic from man-in-the-middle attacks and eavesdropping. 

Docker offers an easy way to encapsulate the OpenVPN server. In this way, the OpenVPN server process and data configuration can be managed more easily at Docker's platforms. The image of the Docker OpenVPN is prebuilt. It contains everything necessary to run the server in a well-balanced and persistent environment. Docker includes scripts that considerably automate the standard use case; however, it also offers complete manual configuration if desired. A Docker volume container is utilized to hold the EasyRSA PKI and also configuration certificate data.
\subsection{ARP spoofing}
ARP (Address Resolution Protocol) spoofing is a kind of security attack in which an attacker sends fake ARP messages over a LAN (local area network). In such attacks, the attacker can link his/her MAC address with the IP address of a legitimate system on the network. As the attacker's MAC address is linked to a legitimate IP address, the attacker will get every bit of data from that specific IP address. 

Initially, Docker developers paid less attention to the fact that the so‐called ARP is employed to map IPv4 to Ethernet hardware (MAC) addresses, which the virtual bridge can also utilize to distribute the Ethernet frames to the correct container as ARP packets are not filtered, so there is no security mechanism available in ARP itself. Therefore containers can indeed imitate other containers or even the host. Such situations can present an ARP spoofing or ARP cache poisoning attack scenario. The NDP (Neighbor Discovery Protocol) in IPv6 is used similarly. 

If an attacker gains access to one of the containers by compromising the container's security, the attacker can obtain, manipulate or redirect any bridge information. This information can be any traffic running among containers and the outside world. An attacker might sniff any secret details (passwords) sent between web applications and database containers in such situations. Moreover, the attacker will also be capable of injecting a malicious payload into network connections.

There are several ways to mitigate such security attacks; one of the most powerful ways to save containers is to run the container without NET\_RAW capability. In this way, programs inside the container will not be able to create PF\_PACKET sockets. Without PF\_PACKET sockets, an ARP spoofing attack cannot be performed. This method has few drawbacks.

Another more suitable way to protect the Docker container from such attacks; is to utilize "ebtables" to filter out Ethernet frames. In this way, ARP packets with the wrong sender protocol or hardware address (ARP spoofing) can be caught and detected [16]. It also allows filtering out the incorrect source addresses (MAC spoofing). In this situation, the attacker has no chance to perform an ARP spoofing attack. 

\section{PROPOSED SOLUTION}
\subsection{Access Control Policy Modules}
We have proposed a solution that is simple and more reliable. This is based on access control methodology to ensure appropriate access management. It is based on specific SELinux types of containerized procedures in a more precise way for the client. In this method, image maintainers ship the SELinux policy module and their images to the host platform. Here, the module will be placed on the host system and outline the types linked with the processes in the Docker image. SELinux modules and the aforementioned Policy Modules (PM) have to fulfill the properties without posing a threat to the host system [17]. The policy modules for an image will be stated in the Docker file and placed in the image metadata at their build-time. For running containerized processes along with some SELinux types, the image-maintainer can label the binaries in the image by particular types and write a type transition procedure. Thus, when the binary is executed, the process is allocated to the SELinux type stated in the rule. It is possible to run different SELinux labels, even if multiple processes have been running at similar the same image. When a Docker container holds different images, all the policy modules for the images that comprise the container will be installed. It also offers SELinux types intended for processes in the parent images. 

The Docker Hub Registry has to ensure that policy modules must not alter the system policy and can only influence processes and resources linked with the DPM itself. Fig. 4 shows that two Docker containers (apache and MySQL) are running, using explicit SELinux types stated in the policy modules inserted in the Docker images. 

It also needs to ensure that new types stated in policy modules have continuously operated inside the boundaries defined through the svirt\_lxc\_net\_t type. A policy model offers the flexibility of outlining numerous types with diverse privileges. Consequently, a container can switch to the least privileged domain required to obtain the present task.

\begin{figure*}[h]
\centerline{\includegraphics[width=8cm, height=4.5cm]{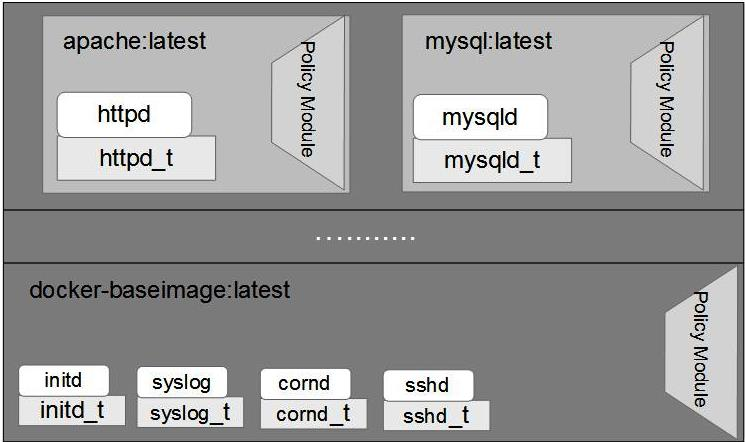}}
\caption{Two Docker containers running processes in, using SELinux types based Policy Modules}
\label{fig1}
\end{figure*}   

\subsection{Secure Deployment Guidelines}
Taking precautionary security measures can save us from huge vulnerabilities. The same case is with the Docker container security. More secure and reliable applications could be developed by taking several precautionary measures for Docker container security. This section will outline some secure deployment guidelines for the Docker platform.

\subsubsection{Docker Images}
As discussed earlier that poisoned images are one of the critical security issues at Docker. However, Docker 1.3 and onward versions offer cryptographic signatures support. Through this approach, a user will discover the actual origin and integrity of official repository images. This capability will reduce the dangers of poisoned images and also reduces the chances of possible security threats. It is highly recommended that all images should be downloaded from the authenticated source and support cryptographic signatures. 

\subsubsection{Network Namespaces} 
Running Docker on a TCP port can cause a serious security hazard for containers. Such an approach permits anyone to get access to a specific port to obtain access to a container. This leads to getting root access on the host or maybe to the Docker group. Therefore, it is critical to ensure that communications are adequately encrypted using SSL while offering access to the daemon over TCP. This approach will prevent unauthorized parties from interacting with it. 

For more enhanced security management, kernel firewall iptables rules can be implemented to docker0. For example, the source IP range can be restricted for a Docker container [18]. This will prevent the container from talking with the outside world. The following iptables filter is used to prevent such access. 
\begin{code}
iptables -t filter -A FORWARD -s 
<source_ip_range> -j REJECT --reject-with 
icmp-admin-prohibited
\end{code}
\subsubsection{Logging and Auditing }
Logging and auditing offer an additional layer of security for Docker security management. In this way, a user can monitor the traffic to ensure that no suspicious activities are being performed.
The following command can be used to access log files outside the container from the host:
\begin{code}
docker run -v /dev/log:/dev/log <container_name> /bin/sh
\end{code}
Log files can also be accessed by using the built-in Docker command:
\begin{code}
docker logs ... (-f to follow log output) 
\end{code}
For permanent storage into a tarball, log files can be exported using the following command:
\begin{code}
docker export ...
\end{code}
\subsubsection{SELinux / AppArmor}

Docker offers Linux kernel security modules like AppArmor and Security-Enhanced Linux (SELinux). These Linux-based Linux kernel security modules can be configured through access control security policies. By configuring these security modules, the users can implement mandatory access controls (MAC) to limit system resources or privileges. 

Through configuring SELinux, Docker will have an additional layer of security through permission checking policy MAC. SELinux manages everything through labels. In the Docker system, every process, file/directory, and system object has a label. The system administrator is using these labels to write rules to manage access between system objects and processes. 

Similar to SELinux, AppArmor is another MAC-based security enhancement model to Linux. AppArmor offers control access to individual programs. Through this model, the administrator can load the security profile into every individual program to restrict and manage the program's capabilities. 

These features are available in Docker version 1.3 and onwards. Docker offers an interface for loading AppArmor's pre-defined profile while launching a new container on AppArmor-supported systems  [19]. 
To load SELinux or AppArmor, security policies are using label confinement; intended for the container. It can be configured using the newly added --security-opt argument in Docker as shown below:  

\begin{code}
--security-opt="label:user:USER": Set the label user for the container
--security-opt="label:role:ROLE": Set the label role for the container
--security-opt="label:type:TYPE": Set the label type for the container
--security-opt="label:level:LEVEL": Set the label level for the container
--security-opt="apparmor:PROFILE": Set the apparmor profile to be applied to the container 

Example: docker run --security-opt=label:level:s0:c100,c200 -i -t centos bash
\end{code}

\subsubsection{Daemon Privileges}
It is recommended not to use the --privileged command because --privileged command will permit the container to access all devices on the host as well as it would provide the container with explicit LSM (i.e., AppArmor or SELinux) configuration. LSM configuration would give a similar level of control as host processes [18]. 

Through avoiding --privileged command could help to diminish the security risks and host compromises. A legitimate user should have the ability to launch the daemon by using the -u option. It can reduce the privileges which are enforced inside the container. For example:

\begin{code}
docker run -u <username> -it <container_name> /bin/bash
\end{code}

\subsubsection{cgroups}
The cgroups, or control groups, offer a way for accounting and limiting the resources for every container. So cgroups offered a great deal of capability to avoid the Denial of Service (DoS) attacks through restricting system resource exhaustion [20]. 

CPU usage:
\begin{code}
docker run -it --rm --cpuset=0,1 -c 2 ... 
\end{code}

Memory usage:
\begin{code}
docker run -it --rm -m 128m ... 
\end{code}

Storage usage:
\begin{code}
docker -d --storage-opt dm.basesize=5G 
\end{code}

\subsubsection{SUID/GUID binaries}
Buffer overflow security attacks can be serious for containers. To avoid such attacks, SUID and SGID binaries should be prohibited. This can be achieved by decreasing the capabilities offered to containers by specific command line arguments.
\begin{code}
docker run -it --rm --cap-drop SETUID --cap-drop SETGID ... 
\end{code}
Another way is to mount the filesystem with the nosuid attribute. By applying this command, a user can avoid SUID resultant buffer overflow security attacks. 

\subsubsection{Devices control group (/dev/*)}
Device isolation is also one of the key ways to avoid several security vulnerabilities. By default, containers include all permissions. To avoid such issues, there is a need that devices should be mount employing the "--device" option, which is built-in, and do not use "-v" with the "--privileged" argument [21]. 

A set of strict permissions can be utilized for the device employing the third set of options, "rwm" to override read, write, and mknod permissions, respectively. For example, sound card read-only permission can be set by the following command: 
\begin{code}
docker run --device=/dev/snd:/dev/snd:r ... 
\end{code}

\subsubsection{Services and Applications}
If a Docker container security is compromised while several sensitive services are running, such situations can lead to a huge disaster. So, to avoid such situation, consider isolating sensitive services. We can add a security layer into our system by running sensitive services (SSH service) on a bastion host or VM. Also, untrusted applications should be avoided to run with root privileges within containers.

\subsubsection{Linux Kernel }           
Sometimes outdated kernels are more likely to be exposed to publicly disclosed vulnerabilities. Therefore, the kernel must be up-to-date with updated utility offered by the system (e.g., apt-get, yum, etc.). There is a great deal of adequate security against memory corruption bugs using a kernel with GRSEC or PAX.

\subsubsection{User Namespaces     }
Currently, user namespaces are not directly supported by Docker. However, they can be used by Docker containers on supported kernels through applying the clone syscall or using the ‘unshare’ feature. UID mapping is presently supported through the LXC driver, however; not in the native libcontainer library. 

User namespaces feature would permit the Docker daemon to execute as an unprivileged user on the host [22]. However, this Docker daemon will appear as executing like root inside containers. 

\subsubsection{libseccomp (and seccomp-bpf extension)       } 
Syscall processes are not significant to system operation. It should be restricted in order to avoid abuse or misuse inside a compromised container. To restrict the Linux kernel’s syscall procedures, the libseccomp library is used. This feature is presently under development. These features are available in LXC driver but not in libcontainer. 

Docker daemon can be restarted to use the LXC driver by using the following command:
\begin{code}
docker -d -e lxc 
\end{code}

\subsubsection{Full Virtualization  } 
Escalation from the container to the host can be so dangerous. It can happen if kernel vulnerability is exploited inside the Docker image. To prevent such issues, use complete virtualization solutions that contain Docker, for example, KVM. Docker offers the capability to nest Docker images to offer a KVM virtualization layer. (Docker-in-Docker utility)

\subsubsection{Security Audits        }
Security audits are one of the key ways to protect the system from significant security risks. Host systems and containers should be audited regularly to assess and identify any vulnerabilities and misconfigurations, if any. These vulnerabilities and misconfigurations could become dangerous for our systems and may compromise intellectual resources [23].

\subsubsection{Multi-tenancy Environments        }
Containers should run on dedicated hosts. It is essential for the security of containers. It becomes more important when the user is dealing with some sensitive operations. 

It is recommended because of the shared nature of Docker containers’ kernel. Therefore, multi-tenancy environments of Docker containers’ kernel can offer secure separation of duty. So, it is highly suggested that containers should be running on dedicated hosts [24]. 
A more secure environment can also be achieved by reducing inter-container communications to a minimal level. It can be achieved by setting the Docker daemon to utilize --icc=false. Also, it is useful to specify -link with Docker run when required. 

\subsubsection{Docker Content Trust }
Content Trust is a new feature that can offer an additional layer of security for containers. It is open-source software that can offer legitimacy of container images. This feature is added in Docker version 1.8.0. It will allow Docker users to confirm the legitimacy of container images (available at any public Docker Hub) before downloading Docker images [25]. 

The basic idea behind this feature is to secure the Docker platform and assure users that they will not be deploying anything possibly hazardous atop their technology infrastructure.

\section{RESULTS AND DISCUSSION  } 
The proposed guidelines offer a great deal of capability to protect against malicious clients or network attacks. Adopting the above-mentioned methodologies and guidelines will ensure a great deal of capability to protect systems against possible breaches. The proposed SELinux types-based Policy Modules for access control will ensure better security and improved access management of possible insider and outsider attacks. Application of these policy modules for different images will offer a comprehensive firewall.  

To perform a network defensive drill, we have to know the difference between real-world attacks and simulation efficiency. Therefore, to test the proposed technique, a DDoS program was executed on the Docker container. Cyber-attacks programs will produce a considerable number of subroutines that are sending requests to attack the targets. Docker containers equipped with the aforementioned tools and capabilities defended better as compared to naive Docker containers. Table 1. shows the experiment aspects where we run Docker version 1.2 on a native OS, with policy modules and specified guidelines. It can be seen that the source to the base presence of policy modules can ensure higher protection and security.  

\begin{table*}[ht]
	\renewcommand{\arraystretch}{1.3}
	\centering
	\caption{Policy Module Application.}
	\begin{tabular}{*{7}{c}}
		\hline
		\textbf{}    & \textbf{Target to Base}   & \textbf{Target to Policy Module}  \\ \hline
		Source to Base           & Threat Present& Partial Threats \\
	    Source to Policy Module  & Partial Threats  &No Threats \\	\hline
	 
	\end{tabular}

\label{tableSeg}
\end{table*}

In March 2015, Docker Inc. published research, which offered a hybrid solution (Combining Containers and Virtual Machines) to enhance security and isolation [26]. Regular Virtual machines cannot be scaled down up to the level of running a single application service. VMs can support a rich set of applications. However, this approach can present some conflicts among collaborating micro-services. On the other hand, running one micro-service per VM is costly from a resource point of view. To resolve these issues, Docker containers can be deployed in conjunction with VMs. This hybrid infrastructure will be based on a combination of containers and VMs. It will allow running an entire group of services in an isolated way and its grouping inside a virtual machine, as shown in Fig.5.

\begin{figure*}[h]
\centerline{\includegraphics[width=16cm, height=11.5cm]{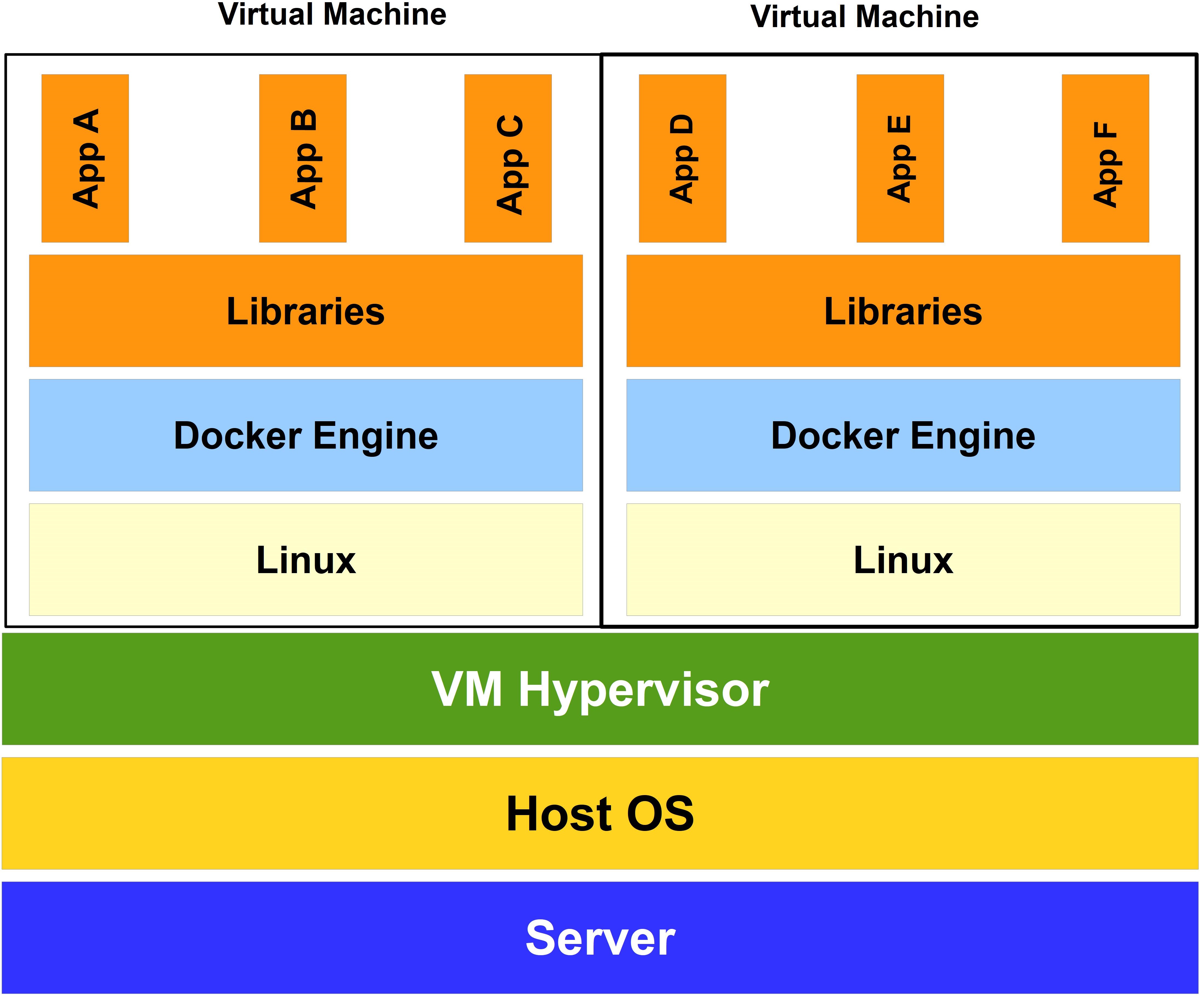}}
\caption{Combining Containers and Virtual Machines}
\label{fig1}
\end{figure*}  
One of the critical features of this approach is the enhanced security by introducing two layers, VMs, and containers, to the distributed applications. The other feature of this technique is to utilize resources in a better way. Moreover, it increases container density, decreasing the number of VMs necessary for the defined isolation and security objectives.

\section{CONCLUSION }
Container applications are getting popular. Docker, LXC, Rocket, or other projects are getting momentum in the container application field. This technology is here to stay. As the technology and related processes mature, they will address many of the risks, a few of which have been outlined above. Docker Containers are offering a lightweight and efficient way to package an application with all its dependencies. However, some security issues hindering their widespread adoption. This research has addressed and outlined some potential security issues and vulnerabilities and offer mitigation strategies to manage these issues. This research also outlines some security deployment strategies to deploy Docker applications more securely and safely. These guidelines and precautionary measures can offer a more secure and reliable container platform for future application development. Currently, Docker has offered its 1.8 version with new updates and fixes. Now Docker is offering an entirely secure container-based application development platform. 

Docker suggests that users can ensure the security of processes by running inside the containers as non-privileged users (i.e., non-root). We also recommend adding a layer of security for our applications by enabling SELinux, AppArmor, GRSEC, or a hardening solution. By configuring the right security policy and following the secure deployment guidelines, we can ensure the greater security of Docker containers.

\section*{References }

1. Moral-García, S., Moral-Rubio, S., Fernández, E.B., Fernández-Medina, E.: “Enterprise security pattern: A model-driven architecture instance”. Computer Standards and Interfaces 36(4), 748-758 (2014). 

2. Kalloniatis, C., Mouratidis, H., Vassilis, M., Islam, S., Gritzalis, S., Kavakli, E.: “Towards the design of secure and privacy-oriented information  systems in the cloud: Identifying the major concepts”. Computer Standards and Interfaces 36(4), 759-775 (2014). 

3. Blanco, C., Rosado, D.G., Sánchez, L.E., Jürjens, J.: “Security in information systems: Advances and new challenges”. Computer Standards and Interfaces 4(36), 687-688 (2014). 

4. Bernstein, D.: Containers and cloud: “From lxc to docker to kubernetes”. IEEE Cloud Computing (3), 81-84 (2014). 

5. Bui, T.: “Analysis of Docker Security”. Aalto University T-110.5291 Seminar on Network Security (2014). 

6. Li, S.-H., Yen, D.C., Chen, S.-C., Chen, P.S., Lu, W.-H.,Cho, C.-C.: “Effects of virtualization on information security”. Computer Standards and Interfaces 42, 1-8 (2015). 

7. Turnbull, J.: “The Docker Book”. ames Turnbull; 1.8.0 edition, (2014)

8. Github: “Docker: the container engine”. https://github.com/docker/docker (2015). Accessed 08 25 2015

9. RedHat: “Get Started with Docker Formatted Container Images on Red Hat Systems”. https://access.redhat.com/articles/881893 (2015). Accessed 07 09 2015

10. Docker: “Announcing docker 1.8: content trust, toolbox, and updates to registry and orchestration”. http://blog.docker.com/2015/08/docker-1-8-content-trust-toolbox-registry-orchestration/ (2015). Accessed 09 02 2015

11. Boettiger, C.: “An introduction to Docker for reproducible research, with examples from the R environment”. arXiv preprint arXiv:1410.0846 (2014). 

12. Merkel, D.: “Docker: lightweight Linux containers for consistent development and deployment”. Linux J. 2014(239), 2 (2014). 

13. Gomes, J., Pina, J., Borges, G., Martins, J., Dias, N., Gomes, H., Manuel, C.: “Exploring Containers for Scientific Computing”. In: 8th Iberian Grid Infrastructure Conference Proceedings, p. 27 

14. Felter, W., Ferreira, A., Rajamony, R., Rubio, J.: “An updated performance comparison of virtual machines and linux containers”. Technology 28, 32 (2014). 

15. Rudenberg, J.: “Docker Image Insecurity”. https://titanous.com/posts/docker-insecurity (2014). Accessed 07 06 2015

16. Nyantec: “Docker networking considered harmful”. https://nyantec.com/en/2015/03/20/docker-networking-considered-harmful/ (2015). Accessed 08 02 2015

17. Mutti, S., Bacis, E., Paraboschi, S.: “Policy specialization to support domain isolation”. In: Proceedings of the 2015 Workshop on Automated Decision Making for Active Cyber Defense 2015, pp. 33-38. ACM

18. Grosperrin, Q.: “Docker Secure Deployment Guidelines”.  (2015). 

19. Docker: “Docker security”. https://docs.docker.com/articles/security/ (2015). Accessed 08 29 2015

20. Goldmann, M.: “Resource management in Docker”. https://goldmann.pl/blog/2014/09/11/resource-management-in-docker/ (2014). Accessed 8 8 2015

21. Vieux, V.: “Announcing Docker 1.2.0”. http://blog.docker.com/2014/08/announcing-docker-1-2-0/ (2014). Accessed 08 12 2015

22. Graber, S.: “Introduction to unprivileged containers” https://www.stgraber.org/2014/01/17/lxc-1-0-unprivileged-containers/ (2014). Accessed 08 22 2015

23. Chou, D.C.: “Cloud computing risk and audit issues”. Computer Standards and Interfaces 42, 137-142 (2015). 

24. Turnbull, J.: “Docker Container Breakout Proof-of-Concept Exploit” (2014). 

25. Firshman, B.: “Announcing docker 1.8: content trust, toolbox, and updates to registry and orchestration”. http://blog.docker.com/2015/08/docker-1-8-content-trust-toolbox-registry-orchestration/ (2015). Accessed 08 01 2015

26. Docker: “Introduction to Container Security”. https://d3oypxn00j2a10.cloudfront.net/assets/img/Docker\%20Security/
WP\_Intro\_to\_container\_security\_03.20.2015.pdf (2015). Accessed 08 09 2015

27. R. Yasrab, J. A. Atkinson, D. M. Wells, A. P. French, T. P. Pridmore, and M. P. Pound, “Rootnav 2.0: Deep
learning for automatic navigation of complex plant root architectures,” GigaScience, vol. 8, no. 11, p. giz123,
2019.

28. R. Yasrab, N. Gu, and X. Zhang, “An encoder-decoder based convolution neural network (cnn) for future advanced
driver assistance system (adas),” Applied Sciences, vol. 7, no. 4, p. 312, 2017.

30. R. Yasrab, J. Zhang, P. Smyth, and M. P. Pound, “Predicting plant growth from time-series data using deep
learning,” Remote Sensing, vol. 13, no. 3, p. 331, 2021.

31. R. Yasrab, “Ecru: An encoder-decoder based convolution neural network (cnn) for road-scene understanding,”
Journal of Imaging, vol. 4, no. 10, p. 116, 2018.

32. R. Yasrab, “Srnet: A shallow skip connection based convolutional neural network design for resolving singularities,”
Journal of Computer Science and Technology, vol. 34, pp. 924–938, 2019.

33. R. Yasrab, W. Jiang, and A. Riaz, “Fighting deepfakes using body language analysis,” Forecasting, vol. 3, no. 2,
pp. 303–321, 2021.

34. R. Yasrab, M. P. Pound, A. P. French, and T. P. Pridmore, “Phenomnet: bridging phenotype-genotype gap: a
cnn-lstm based automatic plant root anatomization system,” bioRxiv, pp. 2020–05, 2020.

35. R. Yasrab, M. P. Pound, A. P. French, and T. P. Pridmore, “Rootnet: A convolutional neural networks for complex
plant root phenotyping from high-definition datasets,” bioRxiv, pp. 2020–05, 2020.

36. R. Yasrab, N. Gu, and Z. Xiaoci, “Dcseg: Decoupled cnn for classification and semantic segmentation,” in
Proceedings of the International Conference on Knowledge and Smart Technologies, 2017.

37. R. Yasrab, N. Gu, and X. Zhang, “Scnet: A simplified encoder-decoder cnn for semantic segmentation,” in 2016
5th International Conference on Computer Science and Network Technology (ICCSNT), pp. 785–789, IEEE, 2016

\end{document}